\documentclass[epsfig,12pt]{article}
\usepackage{amssymb}
\usepackage{amsfonts}
\usepackage{graphicx}
\usepackage{amsmath}

\input epsf.sty
\newtheorem{theorem}{Theorem}
\newtheorem{acknowledgement}[theorem]{Acknowledgement}

\input{tcilatex}
\makeatletter \@addtoreset{equation}{section}
\renewcommand{\theequation}{\thesection.\arabic{equation}}

\begin{document}

\title{%
\rightline{\mbox {\normalsize
{Lab/UFR-HEP0601/GNPHE/0601/VACBT/0601}}\bigskip} \textbf{D-string fluid in
conifold: }\\
\textbf{II. Matrix model for D-droplets on }$\mathbb{S}^{3}$\textbf{\ and }$%
\mathbb{S}^{2}$}
\author{R. Ahl Laamara, L.B Drissi, E H Saidi\thanks{%
h-saidi@fsr.ac.ma} \\
%EndAName
{\small \textit{1.}} {\small \textit{Lab/UFR-Physique des Hautes Energies,
Facult\'{e} des Sciences de Rabat, Morocco.}}\\
{\small \textit{2. Groupement National de Physique des Hautes Energies,
GNPHE; }}\\
{\small \textit{Siege focal, Lab/UFR-HEP, Rabat, Morocco.}}\\
{\small \textit{3. VACBT, Virtual African Centre for Basic Science and
Technology, }}\\
{\small \textit{Focal point Lab/UFR-PHE, Fac Sciences, Rabat, Morocco.}}}
\maketitle

\begin{abstract}
Motivated by similarities between Fractional Quantum Hall (FQH) systems and
aspects of topological string theory on conifold, we continue in the present
paper our previous study (hep-th/0604001, hep-th/0601020) concerning FQH
droplets on conifold. Here we focus our attention on the conifold
sub-varieties $\mathbb{S}^{3}$\textbf{\ }and\textbf{\ }$\mathbb{S}^{2}$ and
study the non commutative quantum dynamics of D1 branes wrapped on a circle.
We give a matrix model proposal for FQH droplets of $N$ point like particles
on $\mathbb{S}^{3}$\textbf{\ }and\textbf{\ }$\mathbb{S}^{2}$ with filling
fraction $\nu =\frac{1}{k}$. We show that the ground state of droplets on $%
\mathbb{S}^{3}$ carries an isospin $j=k\frac{N\left( N-1\right) }{2}$ and
gives remarkably rise to $2j+1$ droplets on $\mathbb{S}^{2}$ with
Cartan-Weyl charge $\left\vert j_{z}\right\vert \leq j$.\bigskip

\textbf{Key words}: \ \ \newline
Fractional quantum Hall droplets, D1 branes, matrix model and non
commutative geometry, conifold, topological string on $T^{\ast }\mathbb{S}%
^{3}$.
\end{abstract}

\tableofcontents

\section{Introduction}

\qquad Recently a matrix model proposal for Laughlin fluid on real 2-sphere $%
\mathbb{S}^{2}$ with a radius $R$ has been studied in $\cite{1}$. Starting
from the field action of a fractional quantum Hall (FQH) particle moving on $%
\mathbb{S}^{2}$ with Kahler potential $K\left( z,\overline{z}\right) =\ln
\left( 1+\overline{z}z\right) $; then promoting the coordinate positions $%
\left\{ z_{a},\overline{z}_{a},\text{ }0\leq a\leq N\right\} $ of the FQH
droplet $N$ particles into $N\times N$ matrix fields $Z_{b}^{a}=Z_{b}^{a}%
\left( t\right) $ and its adjoint conjugate $Z_{a}^{+b}=Z_{a}^{+b}\left(
t\right) $, Morariu and Polychronakos (MP) proposed that the lagrangian
density $\mathcal{L}_{MP}\mathcal{=L}(Z,Z^{+},\mathcal{A})$ describing a
such system reads as,
\begin{eqnarray}
\mathcal{L}_{MP} &=&iB\mathrm{Tr}\left[ R^{2}(1+ZZ^{+})^{-1}\left( \mathcal{D%
}_{t}Z\right) Z^{+}\right]  \notag \\
&&-iB\mathrm{Tr}\left[ R^{2}\left( \mathcal{D}_{t}Z\right)
^{+}(1+ZZ^{+})^{-1}Z\right]  \label{01} \\
&&+\mathrm{Tr}\left[ \Psi ^{+}\left( i\frac{d\Psi }{dt}+\mathcal{A}\Psi
\right) \right] -B\theta \mathrm{Tr}\left( \mathcal{A}\right) ,  \notag
\end{eqnarray}%
with an ordering in which $Z$ and $Z^{\dagger }$ alternate. In this relation
the constant $B$ is the magnitude of the usual background magnetic field,
the $N\times N$ hermitian matrix $\mathcal{A}=\mathcal{A}\left( t\right) $
is the gauge field capturing the fluid incompressibility constraint and $%
\mathcal{D}_{t}Z=\frac{dZ}{dt}-i\left[ \mathcal{A},Z\right] $. The $N$
component vector $\left( \Psi ^{a}\right) $ is the Polychronakos field with
adjoint conjugate $\left( \Psi _{a}^{+}\right) $ and $\theta $ is the non
commutativity parameter related to the magnetic field as $B\theta =k$ with $%
k $ a positive definite integer $\cite{2}$. Recall that the integer $k$
appears in continuous gauge field formulation as the level of non
commutative $U\left( 1\right) $ Chern Simons gauge model and $\nu =\frac{1}{k%
}$ is the filling fraction of Laughlin state; for related studies and
generalizations see also $\cite{3}$-$\cite{90}$.

\qquad In the present paper, we consider an alternative matrix model
approach for FQH droplets on the spheres $\mathbb{S}^{3}$ and $\mathbb{S}%
^{2} $ with filling fraction $\nu $. This study, which is also motivated
from the analysis of dynamics of D-string fluid in conifold $\cite{9}$-$\cite%
{10}$, is based on a constrained method and deals with the spheres as
hypersurfaces in $\mathbb{R}^{4}$. The matrix field action modeling the
system is built as follows:

(\textbf{1}) First, we consider matrix model for FQH droplet on the 3-sphere
$\mathbb{S}^{3}\simeq SU\left( 2\right) $ by using Lagrange method for
dealing with constraint eqs. In this model, the 3-sphere $\mathbb{S}^{3}$ is
thought of as a real hypersurface $\left\vert x\right\vert ^{2}+\left\vert
y\right\vert ^{2}=R^{2}$ embedded in $\mathbb{C}^{2}\simeq \mathbb{R}^{4}$
and the corresponding matrix field action $\mathcal{S}_{3}$ has the
following field dependence $\mathcal{S}_{3}=\mathcal{S}_{3}\left[ X,Y,%
\mathcal{A},\mathcal{C}\right] $ where the gauge field $\mathcal{A}=\mathcal{%
A}\left( t\right) $ captures fluid incompressibility condition and plays the
same role as in eq(\ref{01}). After matrix elevation of the coordinate
positions $\left\{ x_{a}\right\} \rightarrow X$ and $\left\{ y_{a}\right\}
\rightarrow Y$, the resulting matrix hypersurface elevating the 3-sphere
namely,%
\begin{equation}
\mathrm{Tr}\left( X^{+}X+Y^{+}Y\right) =R^{2}N,
\end{equation}%
is treated as a constraint eq and is implemented in the field action via an
extra gauge field $\mathcal{C}=\mathcal{C}\left( t\right) $.

\textbf{(2}) Then starting from the matrix field action $\mathcal{S}_{3}$,
we use the coset realization $SU\left( 2\right) /U_{C}\left( 1\right) $ of $%
\mathbb{S}^{2}\simeq \mathbb{C}\mathbb{P}^{1}$ to deduce the matrix model
proposal for FQH droplet on the 2-sphere. The resulting matrix field action $%
\mathcal{S}_{2}$ has now the matrix field dependence; $\mathcal{S}_{2}=%
\mathcal{S}_{2}\left[ X,Y,\mathcal{A},\mathcal{B},\mathcal{C}\right] $. The
extra gauge field $\mathcal{B}=\mathcal{B}\left( t\right) $ captures the
coset gauge symmetry,
\begin{eqnarray}
X^{\prime }\left( t\right) &=&e^{i\varphi \left( t\right) }X\left( t\right) ,
\notag \\
Y^{\prime }\left( t\right) &=&e^{-i\varphi \left( t\right) }Y\left( t\right)
.
\end{eqnarray}%
In the language of $SU\left( 2\right) $ group generated by the usual
3-operators $\mathbf{J}_{\pm }$ and $\mathbf{J}_{0}$ ( see also eqs(\ref{s0}-%
\ref{s1})), this reduction from $\mathbb{S}^{3}$ down to $\mathbb{S}^{2}$
corresponds just to fixing the charge of the $U_{C}\left( 1\right) $ Cartan
Weyl operator $\mathbf{J}_{0}$. Quantum mechanically, this means that wave
functions $|\Phi >$ describing FQH droplets on $\mathbb{S}^{2}$ should obey,
amongst others, the following constraint eq%
\begin{equation}
\mathbf{J}_{0}|\Phi >=j_{z}|\Phi >,  \label{03}
\end{equation}%
where $j_{z}=mN$ is a relative integer. As an immediate consequence; the
wave functions describing FQH droplets on the 3-sphere are given by $%
SU\left( 2\right) $ highest state representations which we denote as,
\begin{equation}
|\Phi >=\left( \text{ \ }|\Phi _{j,j_{z}}>\text{ \ }\right) ,\qquad
\left\vert j_{z}\right\vert \leq j,  \label{04}
\end{equation}%
with highest weight $j=lN$. This result will be proved later on. The
condition (\ref{03}) corresponds then to picking up one of the basis states
of above isospin $j$ representation. Eqs(\ref{03}-\ref{04}) gives a
remarkable link between FQH systems on the spheres $\mathbb{S}^{3}$ and $%
\mathbb{S}^{2}$. Indeed, a FQH droplet on the 3-sphere carries a isospin $j$
and then is made of $\left( 2j+1\right) $ FQH droplets on $\mathbb{S}^{2}$.

\qquad Note that our approach to FQH droplets on the 2-sphere may be viewed
as a linearized description of MP model. In other words, this way of doing
avoids the high non linearity in the matrix field embodied by the factors $%
(1+ZZ^{+})^{-1}$ and $(1+Z^{+}Z)^{-1}$ of\ eq(\ref{01}) which have infinite
expansion in $ZZ^{+}$ and $Z^{+}Z$ respectively. This aim is achieved by
introducing two extra gauge fields. The new matrix field action $\mathcal{S}%
_{2}\left[ X,Y,\mathcal{A},\mathcal{B},\mathcal{C}\right] =\int dt\mathcal{L}%
_{2}$ with%
\begin{eqnarray}
\mathcal{L}_{2} &=&-iB\mathrm{Tr}\left[ X^{+}\frac{dX}{dt}+Y^{+}\frac{dY}{dt}%
\right] +B\mathrm{Tr}\left[ \mathcal{A}\left( \left[ X^{+},X\right] +\left[
Y^{+},Y\right] -\frac{k}{B}\right) \right]  \notag \\
&&-B\mathrm{Tr}\left[ \mathcal{C}\left( \left( X^{+}X+Y^{+}Y\right) -\frac{l%
}{B}\right) \right] -B\mathrm{Tr}\left[ \mathcal{B}\left( \left(
X^{+}X-Y^{+}Y\right) -\frac{m}{B}\right) \right]  \notag
\end{eqnarray}%
involves three kinds of constraint eqs: (\textbf{i}) The usual fluid
incompressibility condition captured by the gauge field $\mathcal{A}$. (%
\textbf{ii}) The coordinate restriction, describing the embedding $\mathbb{S}%
^{3}$ into $\mathbb{R}^{4}$, and captured by the gauge field $\mathcal{C}$. (%
\textbf{iii}) Reduction of $\mathbb{S}^{3}$ down to $\mathbb{S}^{2}$ using
the standard fibration $\mathbb{S}^{3}\sim \mathbb{S}^{1}\times \mathbb{S}%
^{2}$; the corresponding constraint eq is captured by the gauge field $%
\mathcal{B}$.

An other advantage of our method consists in building FQH extensions based
on geometries going beyond $\mathbb{S}^{2}$. A class of these
generalizations is given by the following special extensions: ($\mathbf{%
\alpha }$) the real 3-sphere $\mathbb{S}^{3}$ with matrix field action $%
\mathcal{S}_{3}=\mathcal{S}_{2}|_{\mathcal{B}=0}$ considered in this paper;
see section 3. ($\mathbf{\beta }$) The K3 complex surface viewed as $T^{\ast
}\mathbb{S}^{2}$ and conifold $T^{\ast }\mathbb{S}^{3}$ under investigation
in $\cite{11}$.

\qquad The presentation of this paper is as follows: In section 2, we
describe FQH droplets on $\mathbb{S}^{3}$ and $\mathbb{S}^{2}$ and derive
the corresponding constraint equations. In section 3, we develop our matrix
model proposal of FQH droplets on the spheres and analyse the structure of
the matrix constraint eqs. In section 4, we consider the quantum version and
build the droplet ground state. Last section is devoted to conclusion and
outlook.

\section{$\mathbb{S}^{3}$ and $\mathbb{S}^{2}$ Conifold sub-varieties in $%
\mathbb{C}^{2}$}

\qquad We start this section by presenting briefly the coordinate frame we
shall use to describe the spheres $\mathbb{S}^{3}$ and $\mathbb{S}^{2}$.
Then we derive the constraint eqs of the system. There are different ways to
parameterize these spheres; but here we will think about them as
hypersurfaces in the complex space $\mathbb{C}^{2}$ $\left( \simeq \mathbb{R}%
^{4}\right) $ with holomorphic coordinates $x$ and $y$. This choice, which
is motivated by the use of manifest isometry, gives also an issue to deal
with possible extensions involving K3 and conifold $\cite{10}$.\newline
In the last part, we describe the classical field actions of particles
moving on the spheres, embedded into $\mathbb{C}^{2}$, in presence of a
constant and strong external magnetic field.

\subsection{FQH constraint eqs on spheres}

\qquad Recall that fluid incompressibility constraint is essential in the
field theoretic modeling of FQH systems. This constraint eq, related to the
fluid approximation, plays a central role in understanding the quantum
dynamics. For FQH systems on non flat manifolds, one should also worry about
curvature effect.

\qquad In the present case, we want to study FQH systems on the spheres $%
\mathbb{S}^{3}$ and $\mathbb{S}^{2}$ embedded in $\mathbb{C}^{2}$. Since
these geometries are non flat, we have then extra constraint relations
besides the usual one dealing with fluid incompressibility. In this
subsection, we want to identify these constraint relations.

\subsubsection{$\mathbb{S}^{3}$ case}

\qquad Starting from the complex two dimension space $\mathbb{C}^{2}$\ with
local holomorphic coordinates $x$ and $y$, we can realize the 3-sphere $%
\mathbb{S}^{3}$ as a real hypersurface $H_{\func{real}}=H_{\func{real}%
}\left( x,y,\overline{x},\overline{y}\right) $ embedded in $\mathbb{C}^{2}$.
In practice this is achieved by restricting the complex coordinates $\left(
x,y\right) $ and their complex conjugates to obey the following constraint
relation,
\begin{equation}
H_{\func{real}}\left( x,y,\overline{x},\overline{y}\right) :\qquad x%
\overline{x}+y\overline{y}=\left\vert x\right\vert ^{2}+\left\vert
y\right\vert ^{2}=R^{2},  \label{b1}
\end{equation}%
where $R$ is the radius of the sphere. For later use note that eq(\ref{b1}),
may be put into various forms. An interesting way is the one using
representations of the $SU\left( 2\right) $ isometry of $\mathbb{S}^{3}$.
For this aim we use the isospinors
\begin{equation}
z^{i}=\left( x,y\right) ,\qquad \overline{z}_{i}=\overline{\left(
z^{i}\right) }=\left( \overline{x},\overline{y}\right) ,\qquad i=1,2.
\end{equation}%
The usual $SU\left( 2\right) $ invariant antisymmetric tensor $\varepsilon
^{ij}$ and its inverse $\varepsilon _{ji}$ raise and lower the $SU\left(
2\right) $ indices. So, eq(\ref{b1}) reads also as follows,%
\begin{equation}
H_{\func{real}}\left( z,z\right) :\qquad \sum_{i=1}^{2}z^{i}\overline{z}%
_{i}=R^{2}.
\end{equation}%
This relation is invariant under the following manifest $SU\left( 2\right) $
rotations $z^{i}=M_{j}^{i}z^{j}$ and $\overline{z}_{i}=\overline{z}%
_{k}M_{i}^{\ast k}$ with $M_{i}^{\ast k}M_{j}^{i}=\delta _{j}^{k}$ and $\det
M=1$. \ We also have the following operators,%
\begin{eqnarray}
\mathrm{J}_{+} &=&z^{i}\frac{\partial }{\partial \overline{z}^{i}},  \notag
\\
\mathrm{J}_{-} &=&\overline{z}^{i}\frac{\partial }{\partial z^{i}},
\label{su} \\
\mathrm{J}_{0} &=&\left( z^{i}\frac{\partial }{\partial z^{i}}-\overline{z}%
^{i}\frac{\partial }{\partial \overline{z}^{i}}\right) ,  \notag
\end{eqnarray}%
satisfying the usual commutation relations of the $SU\left( 2\right) $
algebra,%
\begin{eqnarray}
\left[ \mathrm{J}_{0},\text{\textrm{J}}_{+}\right] &=&2\text{\textrm{J}}_{+},
\notag \\
\left[ \mathrm{J}_{0},\mathrm{J}_{-}\right] &=&-2\mathrm{J}_{-},  \label{su1}
\\
\left[ \mathrm{J}_{+},\mathrm{J}_{-}\right] &=&\mathrm{J}_{0}.  \notag
\end{eqnarray}%
Later on ( see section 4), we shall use creation and annihilation operators
of the quantum matrix model proposal to represent this group symmetry. Now
let us consider a particle with coordinate position $x\left( t\right) ,$ $%
y\left( t\right) $ and $\overline{x}\left( t\right) ,$ $\overline{y}\left(
t\right) $ moving in the complex space $\mathbb{C}^{2}$. The restriction to
a motion on the 3-sphere is achieved by requiring%
\begin{equation}
x\left( t\right) \overline{x}\left( t\right) +y\left( t\right) \overline{y}%
\left( t\right) =R^{2},
\end{equation}%
at every moment t. \newline
For a set of $N$ particles described by $x_{a}\left( t\right) ,$ $%
y_{a}\left( t\right) $ and $\overline{x}_{a}\left( t\right) ,$ $\overline{y}%
_{a}\left( t\right) $ with $a=1,...,N$, moving in the complex space $\mathbb{%
C}^{2}$, the above relation extends as follows,%
\begin{equation}
x_{a}\left( t\right) \overline{x}_{a}\left( t\right) +y_{a}\left( t\right)
\overline{y}_{a}\left( t\right) =R^{2},\qquad a=1,...,N.  \label{b}
\end{equation}%
Besides fluid incompressibility constraint relation, (\ref{b}) represents an
extra condition restricting the dynamics of $N$ particles on $\mathbb{C}^{2}$
down to the 3-sphere. We shall turn to this condition later on; for the
moment let us consider the constraint eqs for the 2-sphere embedded in $%
\mathbb{C}^{2}$ using Hopf fibration of $\mathbb{S}^{3}$.

\subsubsection{$\mathbb{S}^{2}$ as reduced $\mathbb{S}^{3}$}

\qquad There are various ways to parameterize $\mathbb{S}^{2}$; one of these
realizations considered recently in $\cite{1}$ is given by the stereographic
coordinate variables using Kahler potential of $\mathbb{S}^{2}\simeq C%
\mathbb{P}^{1}$. An other interesting coordinate frame, which we shall use
here below, rests on embedding $\mathbb{S}^{2}$ in the complex two space $%
\mathbb{C}^{2}$.

To embed $\mathbb{S}^{2}$ in the complex space $\mathbb{C}^{2}$, one
restricts the complex coordinates $\left( x,y\right) $ to the real
hypersurface (\ref{b1}) and uses the fibration $\mathbb{S}^{3}\simeq \mathbb{%
S}^{1}\times \mathbb{S}^{2}$ to restrict further the 3-sphere down to $%
\mathbb{S}^{2}$. This is achieved by the identification%
\begin{equation}
x^{\prime }\equiv e^{i\varphi }x,\qquad y^{\prime }\equiv e^{-i\varphi }y,
\label{b2}
\end{equation}%
where $\varphi $ is the group parameter of $U_{C}\left( 1\right) $ Cartan
abelian sub-symmetry of the $SU\left( 2\right) $ isometry of the 3-sphere.
From group theoretic view, this identification corresponds to starting from $%
\mathbb{S}^{3}\simeq SU\left( 2\right) $\ and performing the coset,%
\begin{equation}
SU\left( 2\right) \qquad \rightarrow \qquad SU\left( 2\right) /U_{C}\left(
1\right) .
\end{equation}%
Notice that without the identification (\ref{b2}), the hypersurface $H_{%
\func{real}}\left( x,y,\overline{x},\overline{y}\right) $ would describe a $%
\mathbb{S}^{3}$ sphere realized as the fibration of $\mathbb{S}^{1}$ over $%
\mathbb{S}^{2}$. \newline
For a system of $N$ particles with coordinate positions $x_{a}\left(
t\right) ,$ $y_{a}\left( t\right) $ and $\overline{x}_{a}\left( t\right) ,$ $%
\overline{y}_{a}\left( t\right) $, we have in addition to eq(\ref{b}) the
following constraint relation,
\begin{equation}
x_{a}^{\prime }\left( t\right) =e^{i\varphi \left( t\right) }x_{a}\left(
t\right) ,\qquad y_{a}^{\prime }\left( t\right) =e^{-i\varphi \left(
t\right) }y_{a}\left( t\right) ,\qquad a=1,...,N,
\end{equation}%
where the phase $\varphi =\varphi \left( t\right) $\ is time dependent.
Therefore, eqs(\ref{b1}-\ref{b2}) form a set of two constraint relations;
the first one reduces the space $\mathbb{C}^{2}\sim \mathbb{R}^{4}$\ into
the real 3-sphere $\mathbb{S}^{3}$ and the second reduces $\mathbb{S}^{3}$
down to the 2-sphere $\mathbb{S}^{2}$.

\subsection{Classical field action on $\mathbb{S}^{3}$}

\qquad Following $\cite{2}$, the field action $S_{3}=\int dt$ $\mathrm{L}%
_{3} $, describing a particle moving on the real 3-sphere $\mathbb{S}%
^{3}\sim SU\left( 2\right) $ and taken in the condition of FQH systems, is
obtained as follows. Start from the lagrangian density $\mathrm{L}_{4}=%
\mathrm{L}_{4}\left( x,\overline{x},y,\overline{y}\right) $ of a FQH
particle moving on the complex two dimension space $\mathbb{C}^{2}\sim
\mathbb{R}^{4}$ which reads as follows,
\begin{equation}
\mathrm{L}_{4}=-\frac{iB}{2}\left[ \left( \overline{x}\frac{dx}{dt}-x\frac{d%
\overline{x}}{dt}\right) +\left( \overline{y}\frac{dy}{dt}-y\frac{d\overline{%
y}}{dt}\right) \right] ,  \label{b0}
\end{equation}%
which, by using the isodoublet coordinates $z^{i}$, reads also as,%
\begin{equation}
dt\text{ }\mathrm{L}_{4}=-\frac{iB}{2}\left( \overline{z}_{i}dz^{i}-z^{i}d%
\overline{z}_{i}\right) .
\end{equation}%
In eq(\ref{b0}), the field variables $x$ and $y$ are the holomorphic
coordinate variables of $\mathbb{C}^{2}$. To get the lagrangian density $%
\mathrm{L}_{3}$ describing the dynamics of this FQH particle; but now
restricted to moving on $\mathbb{S}^{3}$, one should impose the constraint
eq
\begin{equation}
B\left( x\overline{x}+y\overline{y}\right) =l  \label{ab1}
\end{equation}%
where, because of non commutativity requirement, we have set
\begin{equation}
BR^{2}=l,\qquad R^{2}=\frac{l}{B},
\end{equation}%
with $l$ a positive integer. This condition can be implemented in the above
lagrangian density $\mathrm{L}_{4}$ as usual by using the Lagrange method.
As such the lagrangian density $\mathrm{L}_{3}$ describing the classical
motion of the particle on the 3-sphere reads as,%
\begin{equation}
\mathrm{L}_{3}=-B\left[ \overline{x}\left( i\frac{dx}{dt}+\mathrm{C}x\right)
+\overline{y}\left( i\frac{dy}{dt}+\mathrm{C}y\right) -\mathrm{C}\frac{l}{B}%
\right] .  \label{b31}
\end{equation}%
In this equation, the real field $\mathrm{C}=\mathrm{C}\left( t\right) $ is
the Lagrange gauge field capturing the constraint eq(\ref{ab1}) for
embedding the hypersurface $\mathbb{S}^{3}$ into $\mathbb{C}^{2}$. By
minimizing $\mathrm{L}_{3}$ with respect to the field $\mathrm{C}$, we get
precisely the constraint eq(\ref{ab1}); i.e,%
\begin{equation}
\frac{\partial \mathrm{L}_{3}}{\partial \mathrm{C}}=-B\left( x\overline{x}+y%
\overline{y}\right) +l=0.
\end{equation}%
As we see, eq(\ref{b31}) is an abelian one dimensional gauge theory with
target space $\mathbb{C}^{2}$ and gauge covariant derivatives $\mathcal{D}%
_{t}z=\left( \frac{dz}{dt}-i\mathrm{C}z\right) $ where $z$ stands for $%
x\left( t\right) $ and $y\left( t\right) $. The gauge transformations
leaving $\mathrm{L}_{3}$\ invariant read then as follows,%
\begin{equation}
z^{\prime }=e^{i\lambda }z,\qquad \mathrm{C}^{\prime }=\mathrm{C}+\frac{%
d\lambda }{dt},
\end{equation}%
where $\lambda =\lambda \left( t\right) $ is the gauge group parameter.

\qquad Starting from the above relations, we consider now the classical
system of N FQH particles on $\mathbb{S}^{3}$. Using the $z_{a}^{i}=\left(
x_{a},y_{a}\right) $ and $\overline{z}_{ia}=\left( \overline{x}_{a},%
\overline{y}_{a}\right) $ doublets constrained as,%
\begin{equation}
B\left( x_{a}\overline{x}_{a}+y_{a}\overline{y}_{a}\right) =l,\qquad
a=1,...,N,  \label{b360}
\end{equation}%
the lagrangian density describing the classical dynamics of these particles
on the 3-sphere reads then as
\begin{equation}
\mathrm{L}_{3}\left( z,\overline{z},\mathrm{C}\right) =-B\sum_{a=1}^{N}\left[
i\overline{z}_{ia}\mathcal{D}_{t}z^{ia}-\mathrm{C}_{a}\frac{l}{B}\right] ,
\label{b37}
\end{equation}%
where the gauge covariant derivatives $\mathcal{D}_{t}z^{ia}$ is as before
and where summation over the $SU\left( 2\right) $ index is understood. From
this lagrangian density, the conjugate momenta $\pi _{ia}\left( t\right) $
of the field variable $z^{ia}$ is\ equal to $\pi _{ia}=-iB\overline{z}_{ia}$
and the corresponding Poisson brackets are,
\begin{eqnarray}
\left\{ z^{ia},\overline{z}_{jb}\right\} _{PB} &=&\frac{i}{B}\delta
_{j}^{i}\delta _{b}^{a},  \notag \\
\left\{ z^{ia},z^{jb}\right\} _{PB} &=&\left\{ \overline{z}_{ia},\overline{z}%
_{jb}\right\} _{PB}=0,  \label{b38}
\end{eqnarray}%
where
\begin{equation}
\left\{ F,G\right\} _{PB}=\sum_{l,e}\left( \frac{\partial F}{\partial z^{le}}%
\frac{\partial G}{\partial \pi _{le}}-\frac{\partial F}{\partial \pi _{le}}%
\frac{\partial G}{\partial z^{le}}\right) .
\end{equation}%
In the fluid approximation requiring a large number $N$ of particles, the
previous coordinate position variables $z_{a}^{i}$ and $\overline{z}_{ia}$
are promoted to highest dimensional fields as shown below,%
\begin{equation}
\left\{ z_{a}^{i}\left( t\right) ,\text{ \textrm{C}}_{a}\left( t\right)
\text{, }{\small 1\leq a\leq N}\right\} \qquad \rightarrow \qquad
z^{i}\left( t;\tau ,\overline{\tau },\sigma ,\overline{\sigma }\right) ,%
\text{ \textrm{C}}\left( t;\tau ,\overline{\tau },\sigma ,\overline{\sigma }%
\right) ,  \label{b39}
\end{equation}%
or by using the $x$ and $y$ variables,%
\begin{equation}
\left\{ x_{a}\left( t\right) ,\text{ }y_{a}\left( t\right) \text{, }{\small %
1\leq a\leq N}\right\} \qquad \rightarrow \qquad x\left( t;\tau ,\overline{%
\tau },\sigma ,\overline{\sigma }\right) ,\text{ }y\left( t;\tau ,\overline{%
\tau },\sigma ,\overline{\sigma }\right) .  \label{b40}
\end{equation}%
In this approximation, the constraint relation capturing the
incompressibility condition of the FQH fluid is translated into a condition
on the Jacobian $\mathcal{J}_{ac}$ of the transformation,%
\begin{eqnarray}
\tau \qquad &\rightarrow &\qquad x\left( \tau ,\sigma ;\overline{\tau },%
\overline{\sigma }\right) ,  \notag \\
\sigma \qquad &\rightarrow &\qquad y\left( \tau ,\sigma ;\overline{\tau },%
\overline{\sigma }\right) ,  \label{b41}
\end{eqnarray}%
together with similar transformations for the complex conjugate partners.
Fluid incompressibility requires that the absolute value
\begin{equation}
\left\vert \mathcal{J}_{ac}\right\vert =\left\vert \frac{\partial ^{4}\left(
x,y;\overline{x},\overline{y}\right) }{\partial ^{4}\left( \tau ,\sigma ;%
\overline{\tau },\overline{\sigma }\right) }\right\vert  \label{b42}
\end{equation}%
should satisfy $\left\vert \mathcal{J}_{ac}\right\vert =1$. By expliciting (%
\ref{b42}), we see that $\mathcal{J}_{ac}$ is a polynom $P=P\left( \partial
z^{i},\partial \overline{z}_{i}\right) $ of fourth order in the gradient of
the fields;
\begin{eqnarray}
\mathcal{J}_{ac} &=&\left\{ x,\overline{x}\right\} _{\tau \overline{\tau }%
}\left\{ y,\overline{y}\right\} _{\sigma \overline{\sigma }}+\left\{ x,%
\overline{x}\right\} _{\sigma \overline{\sigma }}\left\{ y,\overline{y}%
\right\} _{\tau \overline{\tau }}  \notag \\
&&+\left\{ x,\overline{x}\right\} _{\sigma \tau }\left\{ y,\overline{y}%
\right\} _{\overline{\tau }\overline{\sigma }}+\left\{ x,\overline{x}%
\right\} _{\overline{\sigma }\overline{\tau }}\left\{ y,\overline{y}\right\}
_{\tau \sigma }  \label{si} \\
&&+\left\{ x,\overline{x}\right\} _{\tau \overline{\sigma }}\left\{ y,%
\overline{y}\right\} _{\overline{\tau }\sigma }+\left\{ x,\overline{x}%
\right\} _{\overline{\tau }\sigma }\left\{ y,\overline{y}\right\} _{\tau
\overline{\sigma }},  \notag
\end{eqnarray}%
where $\left\{ f,g\right\} _{\alpha \beta }$ are defined by $\left( \frac{%
\partial f}{\partial \alpha }\frac{\partial g}{\partial \beta }-\frac{%
\partial f}{\partial \beta }\frac{\partial g}{\partial \alpha }\right) $.
This Jacobian is quadratic in the Poisson brackets and involves the sum of
six terms. To deal with the fluid incompressibility condition $\left\vert
\mathcal{J}_{ac}\right\vert =1$, we begin by simplifying the above
expression by restricting the coordinate transformations (\ref{b41}) to the
sub-class,%
\begin{eqnarray}
\tau \qquad &\rightarrow &\qquad x\left( \tau ,\overline{\tau }\right) ,
\notag \\
\sigma \qquad &\rightarrow &\qquad y\left( \sigma ,\overline{\sigma }\right)
.  \label{b43}
\end{eqnarray}%
With these particular class of general coordinate transformations, eq(\ref%
{si}) simplifies to $\mathcal{J}_{ac}=\left\{ x,\overline{x}\right\} _{\tau
\overline{\tau }}\left\{ y,\overline{y}\right\} _{\sigma \overline{\sigma }}$
and the fluid incompressibility condition reduces to the following relation,%
\begin{equation}
\left\{ x,\overline{x}\right\} _{PB}\left\{ y,\overline{y}\right\}
_{PB}=-\theta ^{2},  \label{b44}
\end{equation}%
where $\theta $ is the usual non commutative coordinates parameters with $%
\theta B=k$. This constraint eq can be solved as $\left\{ x,\overline{x}%
\right\} _{PB}=i\theta $ and $\left\{ y,\overline{y}\right\} _{PB}=i\theta $
or equivalently by using the $SU\left( 2\right) $ doublets as follows,%
\begin{equation}
\left\{ \overline{z}_{i},z^{j}\right\} _{PB}=-i\theta \delta _{i}^{j}.
\label{b45}
\end{equation}%
These special constraint relations constitute then the condition for fluid
incompressibility; they have as usual a non commutative geomerty
interpretation. Implementing, this constraint eq into eqs(\ref{b37}) by
using the Lagrange method, we get the field action $\mathrm{S}_{3}=\mathrm{S}%
_{3}\left( z,\overline{z},\mathrm{C,A}\right) $ for incompressible fluid
running on the 3-sphere,%
\begin{equation}
\mathrm{S}_{3}=-B\int dtd^{2}\tau d^{2}\sigma \left[ i\overline{z}_{i}%
\mathcal{\nabla }_{t}z^{i}-\mathrm{C}\frac{l}{B}-\mathrm{A}\frac{k}{B}\right]
,  \label{b46}
\end{equation}%
where we have set,%
\begin{equation}
\mathcal{\nabla }_{t}z^{i}=\mathcal{D}_{t}z^{i}-i\left\{ \mathrm{A}%
,z^{i}\right\} _{PB},
\end{equation}%
and where $\theta $ is the non commutative parameter considered previously.
The gauge field $\mathrm{A}=\mathrm{A}\left( t,\tau ,\sigma ;\overline{\tau }%
,\overline{\sigma }\right) $ is a real gauge field capturing the fluid
incompressibility constraint equation.

\section{Matrix model of droplets on $\mathbb{S}^{3}$ and $\mathbb{S}^{2}$}

\qquad In this section, we develop a matrix model proposal of fractional
quantum Hall droplets on the real spheres $\mathbb{S}^{3}$ and $\mathbb{S}%
^{2}$. The gauge field theoretical method developed here can be also viewed
as an other way to approach FQH droplets on $\mathbb{S}^{2}$ other than the
one considered recently in $\cite{1}$ by using stereographic coordinates. It
has moreover the property of giving a unified description of FQH systems on $%
\mathbb{S}^{3}$ and $\mathbb{S}^{2}$ and a priori D-string droplets on
conifold which is under investigation in $\cite{11}$; for a Chern-Simons
like description see $\cite{10}$.

\qquad To begin note that using the spheres realizations (\ref{b1},\ref{b2}%
), one can parameterize the FQH droplet $N$ particles moving on $\mathbb{S}%
^{3}$, up to the identification (\ref{b2}) on $\mathbb{S}^{2}$, as follows,%
\begin{eqnarray}
x_{a} &=&x_{a}\left( t\right) ,\qquad \overline{x}_{a}=\overline{x}%
_{a}\left( t\right) ,  \notag \\
y_{a} &=&y_{a}\left( t\right) ,\qquad \overline{y}_{a}=\overline{y}%
_{a}\left( t\right) ,
\end{eqnarray}%
with $a=1,...,N$ indexing the set of particles. Recall that the restriction
of the above FQH particle trajectories on $\mathbb{C}^{2}$ down to the real
three dimension hypersurface $\mathbb{S}^{3}$ is given by the following
constraint relations (\ref{b360}).

\subsection{Matrix field variables}

\qquad In the non commutative $N\times N$ matrix description of the FQH
system on the spheres $\mathbb{S}^{3}$ and $\mathbb{S}^{2}$, one proceeds
more or less as usual. First, one elevates the commuting coordinate position
variables $\left\{ z_{a}^{i}\left( t\right) ,\text{ }\overline{z}_{i}^{a},%
\text{ }1\leq a\leq N\right\} $ of the ambiant space $\mathbb{C}^{2}$ into $%
N\times N$ complex matrices as shown below,
\begin{equation}
X_{0}=\left(
\begin{array}{cccc}
x_{1} & 0 & . & 0 \\
0 & x_{2} & . & 0 \\
. & . & . & . \\
0 & 0 & . & x_{N}%
\end{array}%
\right) ,\qquad Y_{0}=\left(
\begin{array}{cccc}
y_{1} & 0 & . & 0 \\
0 & y_{2} & . & 0 \\
. & . & . & . \\
0 & 0 & . & y_{N}%
\end{array}%
\right)  \label{m0}
\end{equation}%
together with the adjoint matrices $X_{0}^{+}$ and $Y_{0}^{+}$. This is not
the unique way to do, since there are also other equivalent matrices $X$ and
$Y$ related to (\ref{m0}) by using similarity transformations. Second, in
terms of the diagonal matrices (\ref{m0}), the constraint relations (\ref%
{b360}) become,
\begin{equation}
B\left( X_{0}^{+}X_{0}+Y_{0}^{+}Y_{0}\right) =l\text{ }\mathrm{I}_{id}.
\label{c}
\end{equation}%
The hermiticity property of $X^{+}X$ and $Y^{+}Y$ gives the general picture
of eq(\ref{c}) where $X$ and $Y$ are no longer diagonal. This is ensured by
a $U\left( N\right) $ similarity transformation,%
\begin{equation}
P:X_{0}^{+}X_{0}\qquad \rightarrow \qquad X^{+}X=P^{+}\left(
X_{0}^{+}X_{0}\right) P,
\end{equation}%
with the $N\times N$ matrix $P$ satisfies $P^{+}P=PP^{+}=\mathrm{I}_{id}$.
The same may be also done for $Y^{+}Y$. Note in passing that this $U\left(
N\right) $ transformation is in fact a special solution involving adjoint
representation of the $U\left( N\right) $ group; for a more general
description involving the $U\left( N\right) \times U\left( N\right) $ gauge
symmetry as well as bi-fundamental representations, see $\cite{12}$.

\qquad The third step towards the building of the matrix model is to note
that FQH system $\left\{ z_{a}^{i}\left( t\right) ,\text{ }\overline{z}%
_{i}^{a}\left( t\right) ,\text{ }1\leq a\leq N\right\} $ constrained as in
eqs(\ref{b360}) can be described by using $N\times N$ matrices $X$ and $Y$
as well as their adjoints. These matrices encode the coordinate positions of
the FQH droplet and the extra degrees of freedom are eliminated at the level
of the field action by imposing the $U\left( N\right) $ symmetry,%
\begin{eqnarray}
\left( X\text{, }Y\right) \qquad &\rightarrow &\qquad \left( P^{+}XP\text{, }%
P^{+}YP\right)  \notag \\
\left( X^{+}\text{, }Y^{+}\right) \qquad &\rightarrow &\qquad \left(
P^{+}X^{+}P\text{, }P^{+}Y^{+}P\right) .  \label{b10}
\end{eqnarray}%
as well as,
\begin{equation}
B\mathrm{Tr}\left( X^{+}X+Y^{+}Y\right) =lN,  \label{b11}
\end{equation}%
for FQH on the 3-sphere $\mathbb{S}^{3}$. We will show below that the
reduction down to the $\mathbb{S}^{2}$ geometry requires moreover the
condition,%
\begin{equation}
B\mathrm{Tr}\left( X^{+}X-Y^{+}Y\right) =mN,  \label{b12}
\end{equation}%
where $m$ is a relative integer. Combing eqs(\ref{b11}) and (\ref{b12}), one
has%
\begin{equation}
B\mathrm{Tr}\left( Y^{+}Y\right) =\frac{l-m}{2}N
\end{equation}%
and%
\begin{equation}
B\mathrm{Tr}\left( X^{+}X\right) =\frac{l+m}{2}N.
\end{equation}%
Positivity of these quantities leads to the condition,%
\begin{equation}
m\text{ }N\text{ }\leq \text{ }l\text{ }N.
\end{equation}%
This relation recalls the usual $SU\left( 2\right) $ spin projection
inequality,%
\begin{equation}
j_{z}\text{ }\leq \text{ }j.
\end{equation}%
Note that emergence of such $SU\left( 2\right) $ relation is not a strange
thing. We will derive it later on; but for the moment note that this is a
previsible property since the matrix variables X and Y carry a $SU\left(
2\right) $ charge.

\subsection{Lagrangian density}

\qquad The non commutative matrix field lagrangian density describing the
above FQH droplet on the 2-sphere is obtained by following the steps:

(\textbf{a}) Implement the matrix elevation $x\rightarrow X$ and $%
y\rightarrow Y$ into the field action eq(\ref{b0}). This gives a matrix
field action $\mathcal{S}_{4}=\mathcal{S}_{4}\left[ X,Y\right] $ on $\mathbb{%
C}^{2}$.

(\textbf{b}) Embed the constraint eq(\ref{b360}) for droplet dynamics on the
3-sphere by using a Lagrange gauge field $\mathcal{C}=\mathcal{C}\left(
t\right) $. We get then the field action $\mathcal{S}_{3}=\mathcal{S}_{3}%
\left[ X,Y,\mathcal{C}\right] $ on $\mathbb{S}^{3}$

(\textbf{c}) Require invariance of the matrix field action $\mathcal{S}_{3}$
under the $U_{C}\left( 1\right) $ Cartan symmetry (\ref{b2}) to describe
droplet dynamics on the 2-sphere,%
\begin{eqnarray}
X\qquad &\rightarrow &\qquad e^{i\varphi }X,  \notag \\
Y\qquad &\rightarrow &\qquad e^{-i\varphi }Y,
\end{eqnarray}%
This lead to the introduction of gauge field $\mathcal{B}=\mathcal{B}\left(
t\right) $ and the resulting action is $\mathcal{S}_{2}^{\prime }=\mathcal{S}%
_{2}^{\prime }\left[ X,Y,\mathcal{C},\mathcal{B}\right] $

(\textbf{d}) Inject the fluid incompressibility condition by help of a third
Lagrange gauge field $\mathcal{A}=\mathcal{A}\left( t\right) $. The field
action is denoted as $\mathcal{S}_{2}=\mathcal{S}_{2}\left[ X,Y,\mathcal{C},%
\mathcal{B},\mathcal{A}\right] $.

The solution for $\mathcal{S}_{2}=\int dt\mathcal{L}_{2}$ reads, up to a
total derivative, as follows ,%
\begin{eqnarray}
\mathcal{L}_{2} &=&-iB\mathrm{Tr}\left[ X^{+}\frac{dX}{dt}+Y^{+}\frac{dY}{dt}%
\right]  \notag \\
&&+B\mathrm{Tr}\left[ \mathcal{A}\left( \left[ X^{+},X\right] +\left[ Y^{+},Y%
\right] -\frac{k}{B}\right) \right]  \notag \\
&&-B\mathrm{Tr}\left[ \mathcal{C}\left( \left( X^{+}X+Y^{+}Y\right) -\frac{l%
}{B}\right) \right]  \label{d1} \\
&&-B\mathrm{Tr}\left[ \mathcal{B}\left( \left( X^{+}X-Y^{+}Y\right) -\frac{m%
}{B}\right) \right] .  \notag
\end{eqnarray}%
It is obvious that we could recover the matrix model for FQH droplets on the
3-sphere from the above lagrangian density just by setting $\mathcal{B}=0$.
It coincides with the matrix elevation of eq(\ref{b46}).

\qquad Moreover minimizing the matrix field lagrangian density $\mathcal{L}%
_{2}$ with respect to the gauge fields $\mathcal{A}$, $\mathcal{C}$ and $%
\mathcal{B}$, one gets the three following matrix constraint relations,%
\begin{eqnarray}
B\left[ X^{+},X\right] +B\left[ Y^{+},Y\right] &=&k\text{\textrm{\ }}\mathrm{%
I}_{id},  \notag \\
B\left( X^{+}X+Y^{+}Y\right) &=&l\text{ }\mathrm{I}_{id},  \label{ccr} \\
B\left( X^{+}X-Y^{+}Y\right) &=&m\text{ }\mathrm{I}_{id}.  \notag
\end{eqnarray}%
The first relation is the standard one with positive integer $k$; it
captures fluid droplet incompressibility. The second constraint eq restricts
the target space geometry from $\mathbb{C}^{2}$ down to $\mathbb{S}^{3}$; it
describes amongst others the quantization of the radius $R$ of the three
sphere into $\sqrt{\frac{l}{B}}$ units,%
\begin{equation}
R=\sqrt{\frac{l}{B}},\qquad l\in \mathbb{N}^{\ast }.
\end{equation}%
The third condition restricts further $\mathbb{S}^{3}$ down to $\mathbb{S}%
^{2}$; we will give details on this constraint relation when we consider the
quantum matrix model. In that case, the condition gets an remarkable group
theoretic interpretation.

\qquad Notice that like in the study of matrix model for FQH on plane, here
also we have a problem when taking the trace of both sides of the first
relation of eqs(\ref{ccr}). To overcome this difficulty, we extend the
Polychronakos trick by modifying the original set of matrix variables. This
is achieved by adding the Polychronakos field $\Psi =\left( \Psi ^{a}\left(
t\right) \right) $ transforming in the fundamental representation of $%
U\left( N\right) $ with gauge invariant lagrangian density,
\begin{equation}
\sum_{a,b=1}^{N}\Psi _{a}^{+}\left( \delta _{b}^{a}i\frac{d\Psi ^{b}}{dt}+%
\mathcal{A}_{b}^{a}\Psi ^{b}\right) .
\end{equation}%
This field transforms as a scalar field under the isometry of the spheres;
so it has no coupling with the extra gauge fields $\mathcal{B}$ and $%
\mathcal{C}$. The previous matrix field action $\mathcal{S}_{2}$ becomes
then $\mathcal{S}=\int dt\mathcal{L}$,%
\begin{equation}
\mathcal{L}=\mathcal{L}_{2}-\mathrm{Tr}\left[ \Psi ^{+}\left( i\frac{d\Psi }{%
dt}+\mathcal{A}\Psi \right) \right] .  \label{d2}
\end{equation}%
This is the full classical lagrangian density of the matrix model proposal
for FQH droplet on the 2-sphere. Now minimizing (\ref{d2}) with respect to
the gauge fields, we get the following modified constraint relations,%
\begin{eqnarray}
T_{b}^{a} &=&B\left[ X^{+},X\right] _{a}^{b}+B\left[ Y^{+},Y\right]
_{a}^{b}+\Psi _{b}^{+}\Psi ^{a},  \notag \\
E_{0} &=&B\mathrm{Tr}\left( X^{+}X+Y^{+}Y\right) ,  \label{cc} \\
I_{0} &=&B\mathrm{Tr}\left( X^{+}X-Y^{+}Y\right) .  \notag
\end{eqnarray}%
Setting $T_{0}=\sum_{a=1}^{N}\Psi _{a}^{+}\Psi ^{a}$, the above classical
constraint eqs can be also decomposed as follows,%
\begin{eqnarray}
S_{b}^{a} &=&T_{b}^{a}-T_{0}\delta _{b}^{a}=0,  \notag \\
T_{0} &=&k\text{ }N,  \notag \\
E_{0} &=&l\text{ }N,  \label{b17} \\
I_{0} &=&m\text{ }N.  \notag
\end{eqnarray}%
For later use, we need also the conjugate momenta $\Pi _{a}^{b},$ $\Gamma
_{a}^{b}$ and $\Upsilon _{a}$ of the holomorphic matrix field variables $%
X_{b}^{a}$, $Y_{b}^{a}$ and $\Psi ^{a}$ respectively. The appropriate
variations of the lagrangian density (\ref{d1}-\ref{d2}) lead to,
\begin{eqnarray}
\Pi _{a}^{b} &=&-iBX_{a}^{\ast b},  \notag \\
\Gamma _{a}^{b} &=&-iBY_{a}^{\ast b}, \\
\Upsilon _{a} &=&-i\Psi _{a}^{\ast }.  \notag
\end{eqnarray}%
Using the following Poisson bracket for functions F and G,%
\begin{eqnarray}
\left\{ F,G\right\} _{PB} &=&+i\left( \frac{\partial F}{\partial \Psi
_{c}^{\ast }}\frac{\partial G}{\partial \Psi ^{c}}-\frac{\partial F}{%
\partial \Psi ^{c}}\frac{\partial G}{\partial \Psi _{c}^{\ast }}\right)
\notag \\
&&+\frac{i}{B}\left( \frac{\partial F}{\partial X_{b}^{\ast a}}\frac{%
\partial G}{\partial X_{a}^{b}}-\frac{\partial F}{\partial X_{a}^{b}}\frac{%
\partial G}{\partial X_{b}^{\ast a}}\right) \\
&&+\frac{i}{B}\left( \frac{\partial F}{\partial Y_{b}^{\ast a}}\frac{%
\partial G}{\partial Y_{a}^{b}}-\frac{\partial F}{\partial Y_{a}^{b}}\frac{%
\partial G}{\partial Y_{b}^{\ast a}}\right) ,  \notag
\end{eqnarray}%
we have for the canonical variables,%
\begin{eqnarray}
\left\{ \Psi _{a}^{\ast },\Psi ^{b}\right\} &=&+i\delta _{a}^{b},,  \notag \\
\left\{ X_{a}^{\ast d},X_{c}^{b}\right\} &=&+\frac{i}{B}\delta
_{a}^{b}\delta _{c}^{d},  \label{pb} \\
\left\{ Y_{a}^{\ast d},Y_{c}^{b}\right\} &=&+\frac{i}{B}\delta
_{a}^{b}\delta _{c}^{d},  \notag
\end{eqnarray}%
and zero for all remaining others. \newline
With these tools at hand, we turn now to study the quantization of the
matrix model proposal.

\section{Droplet ground state}

\qquad In this section, we consider the quantization of above FQH systems on
the spheres $\mathbb{S}^{3}$ and $\mathbb{S}^{2}$. We derive the explicit
form of the wave functions of their ground states $|\Phi >$ which solve the
quantum version of the constraint eqs(\ref{b17}). We start by giving useful
tools on the algebra of monomials of creation and annihilation operators.
Then we build the ground state for the FQH droplet on $\mathbb{S}^{3}$ and $%
\mathbb{S}^{2}$ respectively.

\subsection{Droplet monomial creators}

\qquad To build the quantum states of the fluid droplets on the spheres, we
need monomials of creation operators satisfying specific properties. In this
subsection, we describe some useful features of these monomials; they will
be used later.

\subsubsection{Algebra of monomials}

\qquad From the Poisson brackets (\ref{pb}) and correspondence principle of
quantum mechanics, the canonical commutation relations for quantum matrix
model read as follows,%
\begin{eqnarray}
\left[ \Psi _{a}^{\dagger },\Psi ^{b}\right] &=&-\delta _{a}^{b},  \notag \\
\left[ X_{a}^{\dagger c},X_{b}^{d}\right] &=&-\theta \delta _{a}^{d}\delta
_{b}^{c},  \label{c1} \\
\left[ Y_{a}^{\dagger c},Y_{b}^{d}\right] &=&-\theta \delta _{a}^{d}\delta
_{b}^{c},  \notag
\end{eqnarray}%
with $\theta =\frac{1}{B}$. We also have,%
\begin{equation}
\left[ \Psi _{a}^{\dagger },X_{b}^{\dagger c}\right] =\left[ \Psi
_{a}^{\dagger },Y_{b}^{\dagger c}\right] =\left[ Y_{a}^{\dagger
c},X_{b}^{\dagger d}\right] ,  \label{c2}
\end{equation}%
together others type $\left[ \Psi _{a}^{\dagger },\Psi _{b}^{\dagger }\right]
=0$ and so on. Then the quantum version of the classical constraint eqs(\ref%
{b17}) reads for the case of droplets on the 3-sphere as%
\begin{eqnarray}
\mathcal{S}_{b}^{a}|\Phi &>&=0,  \notag \\
\mathcal{T}_{0}|\Phi &>&=kN|\Phi >,  \label{qq} \\
\mathcal{E}_{0}|\Phi &>&=lN|\Phi >.  \notag
\end{eqnarray}%
For the case of FQH droplets on the 2-sphere, we have, in addition to eqs(%
\ref{qq}), the following constraint relation%
\begin{equation}
\mathcal{I}_{0}|\Phi >=mN|\Phi >.  \label{qc}
\end{equation}%
The quantum operators $\mathcal{T}_{0}$, $\mathcal{E}_{0}$ and $\mathcal{I}%
_{0}$, associated with the corresponding classical analogues (\ref{cc}), are
given by,%
\begin{eqnarray}
\mathcal{T}_{0} &=&\mathrm{Tr}\left( \Psi ^{\dagger }\Psi \right) ,  \notag
\\
\mathcal{E}_{0} &=&B\mathrm{Tr}\left( X^{\dagger }X+Y^{\dagger }Y\right) ,
\label{cq} \\
\mathcal{I}_{0} &=&B\mathrm{Tr}\left( X^{\dagger }X-Y^{\dagger }Y\right) .
\notag
\end{eqnarray}%
They depend linearly on the creation operators $\Psi ^{\dagger },$\ $%
X^{\dagger }$ and $Y^{\dagger }$\ and their annihilation partners $\Psi $, $%
X $ and $Y$ satisfying the canonical commutation relations (\ref{c1}-\ref{c2}%
). These are hermitian operators ($\mathcal{T}_{0}^{\dagger }=\mathcal{T}%
_{0},$ $\mathcal{E}_{0}^{\dagger }=\mathcal{E}_{0},$ $\mathcal{I}%
_{0}^{\dagger }=\mathcal{I}_{0}$) verifying%
\begin{equation}
\left[ \mathcal{T}_{0},\mathcal{E}_{0}\right] =0,\qquad \left[ \mathcal{T}%
_{0},\mathcal{I}_{0}\right] =0,\qquad \left[ \mathcal{I}_{0},\mathcal{E}_{0}%
\right] =0,
\end{equation}%
and implying that the wave functions describing droplets on the 2-sphere
should depend on three quantum numbers as shown below,%
\begin{equation}
|\Phi >\equiv |\Phi _{k,l,m}>,
\end{equation}%
or more precisely $|\Phi >\equiv |\Phi _{k,j,j_{z}}>$ with the integers $j$
and $j_{z}$ as in introduction section. To get the wave function for the
ground state $|\Phi >$ solution of the quantum constraint eqs(\ref{cq}), it
is interesting to begin by considering the properties of the following
matrix operator monomials,%
\begin{eqnarray}
A_{a_{n}}^{\dagger } &=&\Psi _{b_{0}}^{\dagger }X_{b_{1}}^{\dagger
b_{0}}...X_{a_{n}}^{\dagger b_{n-1}},\qquad n\geq 0,  \notag \\
B_{a_{n}}^{\dagger } &=&\Psi _{b_{0}}^{\dagger }Y_{b_{1}}^{\dagger
b_{0}}...Y_{a_{n}}^{\dagger b_{n-1}},\qquad n\geq 0.  \label{ba}
\end{eqnarray}%
$A_{a_{n}}^{\dagger }$ involves one $\Psi ^{\dagger }$ creator and $\left(
n-1\right) $ creators type $X^{\dagger }$ and $B_{a_{n}}^{\dagger }$ uses
one $\Psi ^{\dagger }$ and $\left( n-1\right) $ operators $Y^{\dagger }$
respectively; they transform in the fundamental of $SU\left( N\right) $. We
also introduce the specific building blocks,%
\begin{equation}
\mathbb{A}_{a_{1}...a_{p}}^{\dagger
}=\dprod\limits_{n=1}^{p}A_{a_{n}}^{\dagger },\qquad \mathbb{B}%
_{a_{1}...a_{q}}^{\dagger }=\dprod\limits_{n=1}^{q}B_{a_{n}}^{\dagger },
\label{ab}
\end{equation}%
more general ones will be given later on.

Concerning the commutation relations of these operators with $\mathcal{S}%
_{b}^{a}$, $\mathcal{T}_{0}$, $\mathcal{E}_{0}$ and $\mathcal{I}_{0}$, it is
useful to arrange them into three classes according to the constraint eq one
is dealing with. Each class consists of two sets; one involving monomials in
$\Psi ^{\dagger }$ and $X^{\dagger }$ and the other uses $\Psi ^{\dagger }$
and $Y^{\dagger }$. The first class concerns commutation relations with the
charge operator $\mathcal{T}_{0}$. We have $\left[ \mathcal{T}%
_{0},X_{b}^{\dagger a}\right] =\left[ \mathcal{T}_{0},Y_{b}^{\dagger a}%
\right] =0$, $\left[ \mathcal{T}_{0},A_{a_{n}}^{\dagger }\right]
=A_{a_{n}}^{\dagger }$, $\left[ \mathcal{T}_{0},B_{a_{n}}^{\dagger }\right]
=B_{a_{n}}^{\dagger }$ and
\begin{eqnarray}
\left[ \mathcal{T}_{0},\mathbb{A}_{a_{1}...a_{p}}^{\dagger }\right] &=&p%
\mathbb{A}_{a_{1}...a_{p}}^{\dagger },  \notag \\
\left[ \mathcal{T}_{0},\mathbb{B}_{a_{1}...a_{q}}^{\dagger }\right] &=&q%
\mathbb{B}_{a_{1}...a_{q}}^{\dagger }.  \label{g1}
\end{eqnarray}%
Similarly, we have $\left[ \mathcal{E}_{0},X_{b}^{\dagger a}\right]
=X_{b}^{\dagger a}$, $\left[ \mathcal{E}_{0},Y_{b}^{\dagger a}\right]
=Y_{b}^{\dagger a}$, $\left[ \mathcal{E}_{0},A_{a_{n}}^{\dagger }\right]
=nA_{a_{n}}^{\dagger }$, $\left[ \mathcal{E}_{0},B_{a_{n}}^{\dagger }\right]
=nB_{a_{n}}^{\dagger }$ and in general,%
\begin{eqnarray}
\left[ \mathcal{E}_{0},\mathbb{A}_{a_{1}...a_{p}}^{\dagger }\right] &=&\frac{%
p\left( p+1\right) }{2}\mathbb{A}_{a_{1}...a_{p}}^{\dagger },  \notag \\
\left[ \mathcal{E}_{0},\mathbb{B}_{a_{1}...a_{q}}^{\dagger }\right] &=&\frac{%
q\left( q+1\right) }{2}\mathbb{B}_{a_{1}...a_{q}}^{\dagger }.
\end{eqnarray}%
Finally, we have the identities $\left[ \mathcal{I}_{0},X_{b}^{\dagger a}%
\right] =X_{b}^{\dagger a},$ $\left[ \mathcal{I}_{0},Y_{b}^{\dagger a}\right]
=-Y_{b}^{\dagger a},$\ $\left[ \mathcal{I}_{0},A_{a_{n}}^{\dagger }\right]
=nA_{a_{n}}^{\dagger }$, $\left[ \mathcal{I}_{0},B_{a_{n}}^{\dagger }\right]
=-nB_{a_{n}}^{\dagger }$ from which we derive,%
\begin{eqnarray}
\left[ \mathcal{I}_{0},\mathbb{A}_{a_{1}...a_{p}}^{\dagger }\right] &=&\frac{%
p\left( p+1\right) }{2}\mathbb{A}_{a_{1}...a_{p}}^{\dagger }  \notag \\
\left[ \mathcal{I}_{0},\mathbb{B}_{a_{1}...a_{q}}^{\dagger }\right] &=&-%
\frac{q\left( q+1\right) }{2}\mathbb{B}_{a_{1}...a_{q}}^{\dagger }.
\label{g5}
\end{eqnarray}%
As one sees, the matrix operators $X^{\dagger }$ and $Y^{\dagger }$ as well
as their adjoint partners play a symmetric role. Thus, the building blocks $%
\mathbb{A}_{a_{1}...a_{q}}^{\dagger }$ and $\mathbb{B}_{a_{1}...a_{q}}^{%
\dagger }$\ should be related by the action under $SU\left( 2\right) $
generators; they are particular condensates of more general ones which we
denote as $\mathbb{D}_{a_{1}...a_{q}}^{\left( n\right) \dagger }$. In what
follows, we study some properties of these objects and their symmetry; then
turn back to the construction of the wave functions.

\subsubsection{$\mathbb{D}_{a_{1}...a_{q}}^{\left( n\right) \dagger }$
building blocks}

\qquad We start by recalling that the classical coordinate variables $x$ and
$y$ are rotated under $SU\left( 2\right) $ isometry of the 3-sphere. This
invariance is generated by the operators $\mathrm{J}_{0}$ and \textrm{J}$%
_{\pm }$ satisfying the commutation relations,%
\begin{eqnarray}
\left[ \mathrm{J}_{0},\text{\textrm{J}}_{+}\right] &=&2\mathrm{J}_{+},
\notag \\
\left[ \mathrm{J}_{0},\mathrm{J}_{-}\right] &=&-2\mathrm{J}_{-},  \label{u}
\\
\left[ \mathrm{J}_{+},\mathrm{J}_{-}\right] &=&\mathrm{J}_{0}.  \notag
\end{eqnarray}%
This symmetry is also valid after the matrix elevation and quantization. To
get the explicit expression of the generators $\mathrm{J}_{0}$ and \textrm{J}%
$_{\pm }$ of this symmetry in quantized matrix model, we should solve
highest weight relations; in particular,%
\begin{eqnarray}
\left[ \mathrm{J}_{0},X\right] &=&-X,\qquad \left[ \mathrm{J}_{0},Y^{\dagger
}\right] =-Y^{\dagger }  \notag \\
\left[ \mathrm{J}_{-},X\right] &=&0,\qquad \left[ \mathrm{J}_{-},Y^{\dagger }%
\right] =0,  \label{j0}
\end{eqnarray}%
which mean that $Y^{\dagger }$ and $X$ carries a spin $\frac{1}{2}$ charge.
We have as well,%
\begin{eqnarray}
\left[ \mathrm{J}_{0},X^{\dagger }\right] &=&X^{\dagger },\qquad \left[
\mathrm{J}_{0},Y\right] =Y  \notag \\
\left[ \mathrm{J}_{+},X^{\dagger }\right] &=&0,\qquad \left[ \mathrm{J}_{+},Y%
\right] =0.  \label{j1}
\end{eqnarray}%
Remembering the properties,%
\begin{equation}
\left[ X^{\dagger },X\right] =-\theta ,\qquad \left[ Y^{\dagger },Y\right]
=-\theta ,
\end{equation}%
and using the relations $\left[ X^{\dagger }X,X\right] =-\theta X$ and $%
\left[ Y^{\dagger }Y,Y\right] =-\theta Y$ as well similar others, eq(\ref{j0}%
) can be solved as follows,%
\begin{equation}
\mathrm{J}_{0}=\frac{1}{\theta }\left( X^{\dagger }X-Y^{\dagger }Y\right) .
\label{s0}
\end{equation}%
This quantum charge operator satisfy $\left[ \mathrm{J}_{0},X^{\dagger }%
\right] =X^{\dagger }$ and $\left[ \mathrm{J}_{0},Y^{\dagger }\right]
=-Y^{\dagger }$. Putting this solution back into eq(\ref{u}), we get%
\begin{equation}
\mathrm{J}_{+}=\frac{1}{\theta }X^{\dagger }Y,\qquad \mathrm{J}_{-}=\frac{1}{%
\theta }Y^{\dagger }X.  \label{s1}
\end{equation}%
From this realization, we can check that this representation obey eq(\ref{u}%
). Moreover, we can check as well that the operators $\mathbb{A}%
_{a_{1}...a_{q}}^{\dagger }$ and $\mathbb{B}_{a_{1}...a_{q}}^{\dagger }$ are
related as
\begin{equation}
\mathbb{B}_{a_{1}...a_{q}}^{\dagger }=\frac{1}{q!}ad^{q}\mathrm{J}_{-}\left(
\mathbb{A}_{a_{1}...a_{q}}^{\dagger }\right) ,\qquad q\geq 1,
\end{equation}%
showing that $\mathbb{A}_{a_{1}...a_{q}}^{\dagger }$ and $\mathbb{B}%
_{a_{1}...a_{q}}^{\dagger }$ are in fact just two special operators of a
more general set of monomials involving both the creators $X^{\dagger }$ and
$Y^{\dagger }$. This set of operators is given by
\begin{equation}
\mathbb{D}_{a_{1}...a_{q}}^{\left( n\right) \dagger }=\frac{\left(
q-n\right) !}{q!}ad^{n}\mathrm{J}_{-}\left( \mathbb{A}_{a_{1}...a_{q}}^{%
\dagger }\right) ,\qquad n=0,1,...,q.  \label{dd}
\end{equation}%
The $\mathbb{D}_{a_{1}...a_{q}}^{\left( n\right) \dagger }$s form then a $%
SU\left( 2\right) $ representation of isospin $j=\frac{q}{2}$. They
constitute the building blocks for the derivation of the quantum matrix
model wave functions. With these tools, we are now in position to build the
wave function for the ground state of the FQH fluids on the spheres.

\subsection{Wave functions for droplets on $\mathbb{S}^{3}$ and $\mathbb{S}%
^{2}$}

\qquad Since $\mathbb{S}^{2}$ is obtained from $\mathbb{S}^{3}$ as a fiber
bundle section, we begin by studying the fundamental wave function for a FQH
fluid droplet on the 3-sphere. Then we consider the computation of the
ground state for the case of FQH droplets on the 2-sphere.

\subsubsection{Ground state for droplets on $\mathbb{S}^{3}$}

\qquad We start by introducing a $U\left( N\right) \times SU\left( 2\right) $
gauge invariant vacuum state $|0>$ describing a state with zero particle.
This state satisfies the symmetries,
\begin{equation}
\mathcal{S}_{b}^{a}|0>=0,\qquad \mathrm{J}_{0,\pm }|0>=0,
\end{equation}%
where $\mathcal{S}_{b}^{a}$ and $\mathrm{J}_{0,\pm }$ are the generators of $%
U\left( N\right) $ and $SU\left( 2\right) $ respectively. This vacum state
is also annihilated by the operators $\Psi ^{b}$, $X_{a}^{b}$\ and $%
Y_{a}^{b} $ of the quantized matrix model; i.e%
\begin{equation}
\Psi ^{b}|0>=0,\qquad X_{a}^{b}|0>=0,\qquad Y_{a}^{b}|0>=0.  \label{e1}
\end{equation}%
By applying monomials of creation operators $\Psi ^{\dagger b}$, $%
X_{a}^{\dagger b}$\ and $Y_{a}^{\dagger b}$, using the properties eqs(\ref%
{g1}-\ref{g5}), and borrowing some ideas from the construction of $\cite{13}$%
, one can build the wave function for FQH droplets on $\mathbb{S}^{3}$
solving the constraint eqs(\ref{qq}). In what follows, we focus our
attention on ground state; the corresponding results are collected in the
following theorem,

\begin{theorem}
\qquad\ \ \newline
\textbf{1}. (\textbf{a}) Creation operators $X_{a}^{\dagger b}$\ and $%
Y_{a}^{\dagger b}$ of quantum matrix model carry isospin charge $\frac{1}{2}$
of the $SU\left( 2\right) $ isometry group of $\mathbb{S}^{3}$. The wave
functions \TEXTsymbol{\vert}$\Phi $\TEXTsymbol{>} built out of these
operators carry then a $SU\left( 2\right) $ isospin charge $j$. \newline
(\textbf{b}) There are $\left[ kN\left( N-1\right) +1\right] $ possible wave
functions describing FQH droplet ground states on the 3-sphere. These ground
states, which solve the constraint eqs(\ref{qq}), are invariant under $%
U\left( N\right) $ gauge symmetry and form a $SU\left( 2\right) $
representation.\newline
\textbf{2}. The ground states $|\Phi _{k,j}>$ of droplets on $\mathbb{S}^{3}$
belong to a highest weight representation of isospin $j=\frac{kN\left(
N-1\right) }{2}=lN$. Besides fluid incompressibility relation $\mathcal{T}%
_{0}|\Phi _{k,j}>=k|\Phi _{k,j}>$, the highest weight states $|\Phi _{k,j}>$
satisfy,%
\begin{eqnarray}
\mathbf{J}_{0}|\Phi _{k,j} &>&=j|\Phi _{k,j}>,  \notag \\
\mathbf{J}_{+}|\Phi _{k,j} &>&=0,\qquad  \label{t1}
\end{eqnarray}%
where $\mathbf{J}_{0}$ and $\mathbf{J}_{\pm }$ are $SU\left( 2\right) $
generators given by eqs(\ref{u}).\newline
\textbf{3}. The quantum number $l$ describing the quantization of the radius
squared of the 3-sphere ($l=BR^{2}$) is given by%
\begin{equation}
l=\frac{k\left( N-1\right) }{2}.  \label{k}
\end{equation}%
The meaning of number $k$ is as usual, however the apparition of the factor $%
\frac{\left( N-1\right) }{2}$ is a consequence of matrix elevation of the
3-sphere geometry. Requiring the $N\times N$ matrices to be in a $SU\left(
2\right) $ isospin $l_{1}$ representation, one discovers by solving the
equation $2l_{1}+1=N$ that $l_{1}=\frac{\left( N-1\right) }{2}$; it
corresponds to setting $k=1$ in eq(\ref{k}).\newline
\textbf{4}. The explicit expression of the highest weight state $|\Phi
_{k,j}>$ describing the ground state of droplets on $\mathbb{S}^{3}$ is
given by%
\begin{equation}
|\Phi _{k,j}>=\left[ \left( \epsilon ^{a_{0}a_{1}...a_{N-1}}\Psi
_{a_{0}}^{\dagger }\mathbb{A}_{a_{1}...a_{N-1}}^{\dagger }\right) ^{k}\right]
|0>,  \label{t2}
\end{equation}%
where $\mathbb{A}_{a_{1}...a_{N-1}}^{\dagger }$ is as in eq(\ref{ba}) and
where $\epsilon ^{a_{0}a_{1}...a_{N-1}}$ is the completely antisymmetric
invariant tensor of $SU\left( N\right) $ gauge symmetry. Excited states are
obtained in a quite similar manner as in $\cite{13}$.
\end{theorem}

The proof of this theorem follows immediately by using the tools given in
previous subsections. Let us give below some indications. Notice that the
first relation of the constraint eqs(\ref{qq}) is solved in same manner as
done in $\cite{13}$. This is built by using $SU\left( N\right) $ invariants,
in particular the trace, determinant and the completely antisymmetric
invariant tensor $\epsilon ^{a_{1}...a_{N}}$ which contract N indices of the
$U\left( N\right) $ gauge group. Focusing on one of the matrix variables;
say the creator $X_{ca}^{\dagger }$ and borrowing results from $\cite{12,13}$%
, the first and second constraint relations of eqs(\ref{qq}) tell us that
\begin{equation}
|\Phi _{k,j,0}>=\left[ \left( \epsilon ^{a_{0}a_{1}...a_{N-1}}\Psi
_{a_{0}}^{\dagger }\mathbb{A}_{a_{1}...a_{N-1}}^{\dagger }\right) ^{k}\right]
|0>  \label{e4}
\end{equation}%
is a candidate for the wave function. \newline
By using the relations $\mathrm{J}_{0}|0>=0$, $\left[ \mathrm{J}_{0},\Psi
_{a_{0}}^{\dagger }\right] =0$ and $\left[ \mathrm{J}_{0},\mathbb{A}%
_{a_{1}...a_{N-1}}^{\dagger }\right] =\frac{N\left( N-1\right) }{2}\mathbb{A}%
_{a_{1}...a_{N-1}}^{\dagger }$, we find that the above quantum state obeys
the following relations,%
\begin{equation}
\mathrm{J}_{0}|\Phi _{k,j}>=k\frac{N\left( N-1\right) }{2}|\Phi _{k,j}>.
\end{equation}%
Similarly using $\left[ \mathrm{J}_{+},X^{\dagger }\right] =0$ and then $%
\left[ \mathrm{J}_{+},\mathbb{A}_{a_{1}...a_{N-1}}^{\dagger }\right] =0$, we
get \ $\mathrm{J}_{+}|\Phi _{k,j}>=0$. These two relations show that $|\Phi
_{k,j}>$ is a $SU\left( 2\right) $ highest weight state with highest weight $%
j$ given by%
\begin{equation}
j=k\frac{N\left( N-1\right) }{2},
\end{equation}%
and should be thought of as the top vector basis%
\begin{equation}
|\Phi _{k,j}>\equiv |\Phi _{k,j,j_{z}}>,\qquad j_{z}=j
\end{equation}%
of a $\left( 2j+1\right) $\ component vector $\left( |\Phi
_{k,j,j_{z}}>\right) $. Using eq(\ref{e4}), one can build the states of the $%
SU\left( 2\right) $ highest weight representation by applying the monomials $%
\left( \mathrm{J}_{-}\right) ^{p}$. We then have%
\begin{equation}
|\Phi _{k,j,j-p}>=\frac{\left( j-p\right) !}{j!}\left( \mathrm{J}_{-}\right)
^{p}|\Phi _{k,j,j}>,  \label{s3}
\end{equation}%
where the integer $p$ takes values as $p=0,...,\frac{kN\left( N-1\right) }{2}
$. Using the explicit expression of $\mathrm{J}_{-}=\frac{1}{\theta }%
Y^{\dagger }X$ as well as the relation $\left[ \mathrm{J}_{-},X^{\dagger }%
\right] =Y^{\dagger }$, it is not difficult to see that the lowest state is
given by%
\begin{equation}
|\Phi _{k,j,-j}>=\left[ \epsilon ^{a_{0}a_{1}...a_{N-1}}\left( \Psi
_{a_{0}}^{\dagger }\mathbb{B}_{a_{1}...a_{N-1}}^{\dagger }\right) ^{k}\right]
|0>,
\end{equation}%
involving only monomials built of the creation operators $Y^{\dagger }$.
From this solution, one can compute the mean values of $\mathcal{T}_{0}$ and
$\mathcal{E}_{0}$ operators. By using the explicit expression of the wave
function and the algebra of the creation and annihilation operators, we find
for the second relation of constraint eqs(\ref{qc})%
\begin{equation}
\frac{<\Phi _{k,j,j-p}|\mathcal{T}_{0}|\Phi _{k,j,j-p}>}{<\Phi
_{k,j,j-p}|\Phi _{k,j,j-p}>}=k\left[ 1+p+\left( N-p-1\right) \right] =kN.
\end{equation}%
Following the same method, we can compute in a similar manner the mean value
$<\Phi |\mathcal{E}_{0}|\Phi >$ from which we determine the integer $l$. We
get precisely $l=k\frac{\left( N-1\right) }{2}$ as in theorem.

\subsubsection{Ground states for droplets on $\mathbb{S}^{2}$}

\qquad The analysis that we have just carried out in the $\mathbb{S}^{3}$
case is also used for building the ground state of FQH droplets on the
2-sphere. In that context, the construction of the low energy wave function
follows directly from the link between the isometries of the two spheres. In
particular, we get it by using the two following:

(\textbf{a}) The fibration of the 3-sphere $\mathbb{S}^{3}$ as a circle $%
\mathbb{S}^{1}$ fibered on a base $\mathbb{S}^{2}$. In the isometry group
theoretic language, this fibration reads as follows,%
\begin{eqnarray}
\mathbb{S}^{3} &\simeq &SU\left( 2\right) ,  \notag \\
\mathbb{S}^{2} &\simeq &SU\left( 2\right) /U_{C}\left( 1\right) , \\
\mathbb{S}^{1} &\simeq &U_{C}\left( 1\right) .  \notag
\end{eqnarray}%
and shows that $\mathbb{S}^{2}$ may be obtained by fixing the abelian $%
U_{C}\left( 1\right) $ gauge subsymmetry of the isometry of the 3-sphere.

(\textbf{b}) Droplets on the 2-sphere are described by $SU\left( 2\right)
/U_{C}\left( 1\right) $ representations. These are obtained from those of$\
SU\left( 2\right) $ representations by projection of the isospin $j$ highest
weight vector on one of the $\left( 2j+1\right) $ possible directions $j_{z}$%
. With this indication, we can solve without difficulty the constraint eqs
of droplets on $\mathbb{S}^{2}$.

\qquad To get the explicit expression of the wave function of the ground
state describing the FQH droplet on the 2-sphere, we start from the solution
(\ref{s3})%
\begin{equation}
\left\{ |\Phi _{k,j,j-p}>\text{, \ \ \ }0\leq p\leq \frac{1}{2}kN\left(
N-1\right) \right\} ,
\end{equation}%
obtained for the 3-sphere eq(\ref{t1}-\ref{t2}). Then, we restrict $\mathbb{S%
}^{3}$ down to $\mathbb{S}^{2}$ geometry by imposing the constraint eq $%
\mathcal{I}_{0}|\Phi >=m|\Phi >$. This constraint eq corresponds just to
singling out one of the state of the highest weight representation. Thus the
wave function with spin projection $m$ reads as follows,%
\begin{equation}
|\Phi _{SU\left( 2\right) /U_{C}\left( 1\right) }>=|\Phi _{k,j,j-m}>,
\end{equation}%
where $m$ is one of the integers belonging to the set $\left[ -l,+l\right] $%
. Using the building block operators $\mathbb{D}_{a_{1}...a_{N-1}}^{\left(
n_{i}\right) \dagger }$ eq(\ref{dd}), we can rewrite the above fundamental
wave function as follows,%
\begin{equation}
|\Phi _{k,j,j-m}>=\left( \dprod\limits_{i=1}^{k}\left[ \epsilon
^{a_{0}a_{1}...a_{N-1}}\Psi _{a_{0}}^{\dagger }\mathbb{D}%
_{a_{1}...a_{N-1}}^{\left( n_{i}\right) \dagger }\right] \right) |0>,
\label{fs2}
\end{equation}%
where $k$ and $l$ are same as in previous theorem and the integer $m$ is
given by,
\begin{equation}
m=\sum_{i=1}^{k}\left( \frac{N\left( N-1\right) }{2}-2n_{i}\right) ,
\end{equation}%
with $n_{i}=0,...,\frac{N\left( N-1\right) }{2}$. \newline
Comparing this result with its $\mathbb{S}^{3}$ analogue, it is illuminating
to recognize that one FQH droplet moving on the 3-sphere manifests as,
\begin{equation}
2\frac{kN\left( N-1\right) }{2}+1=kN\left( N-1\right) +1
\end{equation}%
FQH droplets on $\mathbb{S}^{2}$. As an illustrating example, we consider
below the $\mathbb{S}^{3}$ FQH droplet $j=1$ by taking $k=1$ and $N=2$. It
consists of the three discs representing $\mathbb{S}^{2}$ FQH droplets with
isospin projections $j_{z}=-1,0,+1$; see also figure.
\begin{figure}[tbph]
\begin{center}
\epsfxsize=13cm\epsfbox{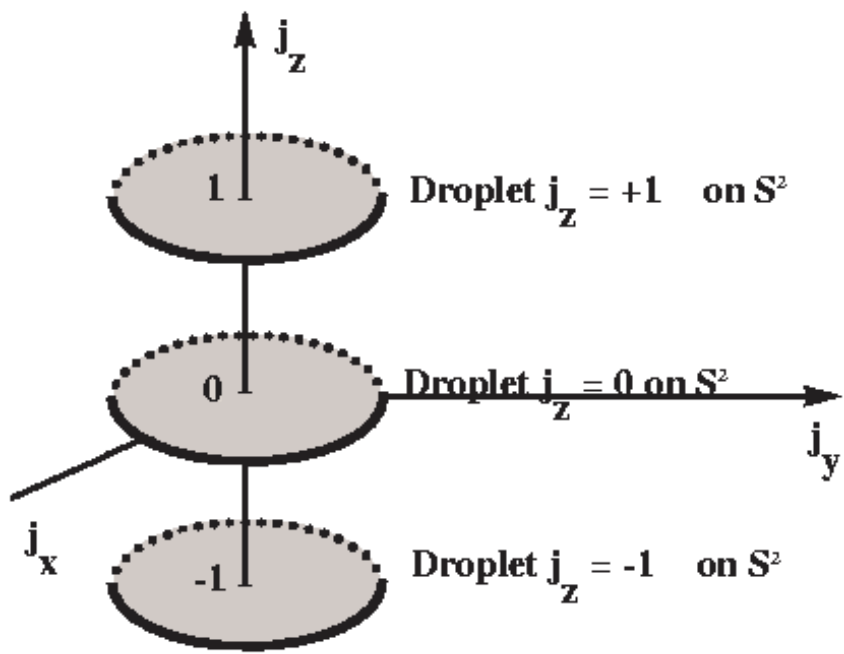}
\end{center}
\caption{This figure gives a schematic illustration of FQH droplet on $%
\mathbb{S}^{3}$ with isospin $j=1$; which is constituted of three FQH
droplets on $\mathbb{S}^{2}$.}
\end{figure}

\section{Conclusion and outlook}

\qquad In this paper, we have developed a matrix model proposal for
fractional quantum Hall droplets on the spheres $\mathbb{S}^{3}\simeq
SU\left( 2\right) $ and $\mathbb{S}^{2}\simeq SU\left( 2\right) /U_{C}\left(
1\right) $. These geometries are realized as hypersurfaces embedded in $%
\mathbb{C}^{2}\simeq \mathbb{R}^{4}$ and treated as constraint relations
implemented in the field action by help of gauge fields. For the case of
droplets on $\mathbb{S}^{2}$, this approach may be viewed as an alternative
method to the highly non linear matrix model proposal considered recently in
$\cite{1}$. Moreover, the wave function eq(\ref{fs2}), derived in this
paper, completes the results of the above mentioned study. An other powerful
point of our proposal is its unified description of both $\mathbb{S}^{3}$
and $\mathbb{S}^{2}$ spheres.

Among the main results obtained in this study, we mention the derivation of
a new matrix model proposal for FQH droplets on $\mathbb{S}^{3}$ and its
reduction down to $\mathbb{S}^{2}$ by borrowing ideas from gauge constrained
systems. After having identified the underlying quantum matrix constraint
eqs(\ref{qc}-\ref{cq}), we have worked out explicitly the droplet ground
state solutions for both spheres $\mathbb{S}^{3}$ and $\mathbb{S}^{2}$. We
have found that isometries of these manifolds play a central role in
building the wave function for ground state of droplets. In particular, we
have shown that FQH droplets on $\mathbb{S}^{3}$ carries an isospin $j$
equal to,
\begin{equation}
j=kN\frac{\left( N-1\right) }{2},\qquad l=kl_{1},\qquad l_{1}=\frac{\left(
N-1\right) }{2},
\end{equation}%
with $\frac{1}{k}$ being the filling fraction of the Laughlin state. We have
also found that a generic droplet on $\mathbb{S}^{3}$ with isospin $j$ and
ground state,%
\begin{equation}
|\Phi _{k,j}>=\left\{ \text{ \ }|\Phi _{k,j,j-p}>,\text{ \ \ }\left\vert
p\right\vert \leq j\text{ \ }\right\} ,
\end{equation}%
is made of $\left( 2j+1\right) $ droplets on the 2-sphere,%
\begin{equation}
|\Phi _{k,j,j-p}>,\qquad p\in \left\{ -j,...,j\right\} ,
\end{equation}%
with fixed value of the relative integer $p$. It would therefore be
interesting to construct matrix model for FQH droplets on group manifolds $G$
and cosets $G/H$; then check whether the FQH ground state property relating $%
\mathbb{S}^{3}\simeq SU\left( 2\right) $ and $\mathbb{S}^{2}\simeq SU\left(
2\right) /U_{C}\left( 1\right) $ is a general feature valid as well for FQH
droplets on higher dimensional group manifolds.

In the end of this paper, we would like to add that the next step of our FQH
project is to push further this method by applying it to other particular
examples. Our immediate interest concerns the derivation of matrix model for
wrapped D-string droplets on $K3\simeq T^{\ast }\mathbb{S}^{2}$ and conifold
$T^{\ast }\mathbb{S}^{3}$ geometries. This is important for our quest in
looking for the connection between FQH droplets on $T^{\ast }\mathbb{S}^{3}$
and the partition function of topological string B model on conifold.
Progress in this direction will be reported elsewhere.

\begin{acknowledgement}
\qquad\ \ \newline
This research work is supported by the program Protars III D12/25, CNRST.
\end{acknowledgement}

\end{document}